\documentclass[%
 reprint,
superscriptaddress,
 amsmath,amssymb,
 aps,
]{revtex4-2}
\usepackage{siunitx}
\usepackage{graphicx}  
\usepackage{subfigure}
\usepackage{multirow}
\usepackage{tabularx}
\usepackage{fancyhdr}
\usepackage{longtable}
\usepackage{parskip}
\usepackage[T1]{fontenc}
\usepackage{dcolumn}   
\usepackage{gensymb}
\usepackage{bm}        
\usepackage{amsfonts}  
\usepackage{amsmath}   
\usepackage{amssymb}   
\usepackage{hhline}
\usepackage{xcolor}
\usepackage{ulem}
\usepackage{gensymb}
\begin{document}

\title{Non-Fermi-liquid behavior in a ferromagnetic heavy fermion system CeTi$_{1-x}$V$_{x}$Ge$_{3}$}

\author{R.-Z. Lin}
\affiliation {\it Department of Physics, National Cheng Kung University, Tainan 701, Taiwan}
\affiliation {\it Center for Quantum Frontiers of Research \& Technology (QFort), National Cheng Kung University, Tainan 701, Taiwan}
\affiliation{\it Taiwan Consortium of Emergent Crystalline Materials, National Science and Technology Council, Taipei 10622, Taiwan}
\author{H. Jin}
\affiliation{\it Department of Physics and Astronomy, University of California, Davis, California 95616, USA}
\author{P. Klavins}
\affiliation{\it Department of Physics and Astronomy, University of California, Davis, California 95616, USA}
\author{W.-T. Chen}
\affiliation {\it Center for Condensed Matter Sciences, National Taiwan University, Taipei 10617, Taiwan}
\affiliation {\it Center of Atomic Initiative for New Materials, National Taiwan University, Taipei 10617, Taiwan}
\affiliation{\it Taiwan Consortium of Emergent Crystalline Materials, National Science and Technology Council, Taipei 10622, Taiwan}
\author{Y.-Y. Chang}
\affiliation{\it Department of Electrophysics, National Yang-Ming Chiao-Tung University, Hsinchu 30010, Taiwan}
\author{C.-H. Chung}
\affiliation{\it Department of Electrophysics, National Yang-Ming Chiao-Tung University, Hsinchu 30010, Taiwan}
\affiliation{\it Physics Division, National Center for Theoretical Sciences, Hsinchu 30013, Taiwan}
\author{V. Taufour}
\affiliation{\it Department of Physics and Astronomy, University of California, Davis, California 95616, USA}
\author{C.-L. Huang}
\email{clh@phys.ncku.edu.tw}
\affiliation {\it Department of Physics, National Cheng Kung University, Tainan 701, Taiwan}
\affiliation {\it Center for Quantum Frontiers of Research \& Technology (QFort), National Cheng Kung University, Tainan 701, Taiwan}
\affiliation{\it Taiwan Consortium of Emergent Crystalline Materials, National Science and Technology Council, Taipei 10622, Taiwan}

\date{\today}

\begin{abstract}  
An investigation of the thermodynamic and electrical transport properties of the isoelectronic chemical substitution series CeTi$_{1-x}$V$_{x}$Ge$_{3}$ (CTVG) single crystals is reported. As $x$ increases, the ferromagnetic (FM) transition temperature is suppressed, reaching absolute zero at the critical concentration $x = 0.4$, where a non-Fermi-liquid low-temperature specific heat and electrical resistivity, as well as the hyperscaling of specific heat and magnetization are found. Our study clearly identifies an FM quantum critical point (QCP) in CTVG. The obtained critical exponents suggest that CTVG falls in the preasymptotic region of the disorder-tuned FM QCP predicted by the Belitz-Kirkpatrick-Vojta theory.

\end{abstract}
\maketitle

The $f$-electron system exhibits diverse magnetic behaviors, which has attracted the attention of many researchers. This diversity in magnetic behavior arises from the competition between the magnetic Ruderman–Kittel–Kasuya–Yosida (RKKY) interaction, the non-magnetic Kondo effect, and the crystal electric field effect. Because the system's ground state is delicately balanced among these three factors, non-thermal tuning parameters, such as magnetic field, pressure, and chemical substitution, are often used to alter the system's underlying physical behavior \cite{Stewart2001,Loehneysen2007,Manuel2016}. Among these, the most eye-catching physical behavior is when long-range magnetic order is continuously suppressed to absolute zero, reaching a magnetic quantum critical point (QCP). 

Depending on the type of magnetic order before tuning, the magnetic QCP can be subdivided into two kinds: antiferromagnetic (AFM) and ferromagnetic (FM) QCPs. The AFM QCP seems easier to obtain than the FM QCP. The reason for this is that when suppressing the FM ordering temperature to zero, in addition to the inherent order parameter fluctuation energy, additional soft modes appear, leading to avoidance of single fixed point \cite{Belitz1999}. The only known ferromagnet that can be tuned to a QCP by hydrostatic pressure is CeRh$_6$Ge$_4$ and its mechanism remains elusive \cite{Shen2020,Kotegawa2019JPSJ}. Another more discernible approach is proposed by Belitz, Kirkpatrick, and Vojta (BKV), suggesting that it is possible to reinstate the FM QCP by introducing a suitable level of quenched disorder \cite{Belitz1999,Sang2014,Kirkpatrick2014,Kirkpatrick2015}. A handful of ferromagnets have confirmed BKV's prediction and are reported to exhibit QCPs, including UCo$_{1-x}$Fe$_{x}$Ge \cite{Huang2016}, (Mn$_{1-x}$Cr$_{x}$)Si \cite{Mishra2020}, 
NiCoCr$_{x}$ \cite{Sales2017}, Ce(Pd$_{1-x}$Ni$_{x}$)$_{2}$P$_{2}$ \cite{Lai2018}, and Ni$_{1-x}$Rh$_{x}$ \cite{Huang2020,Lin2022}. In these studies, researchers observed the logarithmic or power-law divergence of the specific heat coefficient and magnetic susceptibility as the temperature decreases. Some of these studies conducted scaling analyses to compare the critical exponents with those predicted by the BKV theory. The BKV theory also predicts the temperature dependence of the electrical resistivity, which probes how quasiparticles scatter from the order parameter fluctuations. However, studies on resistivity in disorder-tuned ferromagnets are rarely reported. One reason for this scarcity is that most of the research employs polycrystalline samples, where grain boundaries can compromise the data, leading to an ambiguous interpretation of the role of quantum fluctuations. Conducting resistivity measurements on single crystalline disorder-tuned ferromagnets are imperative, as it allows for a more comprehensive examination of the completeness of the BKV theory. 

CeTiGe$_{3}$ has been reported as a rare FM Kondo system with a Curie temperature $T_{\rm C} = 14$~K \cite{Manfrinetti2005}. The FM order can be reduced by the hydrostatic pressure, suggesting that CeTiGe$_{3}$ is located at the right-hand side of the $T_{\rm C}(J)$ maximum in the Doniach model \cite{Doniach1977}. This also implies that the transition is sensitive to chemical substitution, leading to the change in chemical pressure and modification of the band structure. By substituting vanadium for titanium in CeTi$_{1-x}$V$_{x}$Ge$_{3}$ (CTVG), researchers have successfully synthesized this compound in both polycrystalline and single crystalline forms. In both cases, the full suppression of FM order is observed at the critical concentration $x{_{cr}} = 0.4$. \cite{Kittler2013,Jin2022}. However,these reports lack low-temperature experimental evidence that can demonstrate the existence of a FM QCP. In this Letter, we report electrical resistivity, magnetization, and specific heat measurements of CTVG single crystals, with a focus on the $x{_{cr}}$ sample. The results clearly indicate several pieces of evidence for an FM QCP in CTVG, and the obtained quantum critical exponents, as well as the temperature dependence of resistivity, are consistent with the framework of the BKV theory for disorder-tuned FM quantum criticality in the preasymptotic region.

The CTVG single crystals were grown using the self-flux method, with varying values of $x$ including 0, 0.1, 0.3, 0.4, 0.9, and 1.0. The detail of the growth has been described in Ref.~\cite{Jin2022}. For $0.8 >x > 0.4$, we were unable to obtain single crystals, and furthermore, these polycrystals often contain the magnetic impurity CeGe$_{x}$. Therefore, we omit these concentrations. A nonmagnetic reference LaTi$_{0.55}$V$_{0.45}$Ge$_{3}$ was grown using the arc-melting method. Specific-heat measurements employing
thermal-relaxation calorimetry and the 4-point-contact electrical resistivity were performed for current flow in the $a-b$ plane in the temperature range from 0.05 to 300~K using a Dynacool physical properties measurement system
(Dynacool PPMS, Quantum Design) equipped with a 9 T magnet. Magnetization measurements were carried out in a SQUID magnetometer (MPMS, Quantum Design) in the temperature range from 1.8~K to 300~K and in external fields up to 7~T. For all measurements the magnetic field is applied parallel to the hexagonal $c-$axis.

\begin{figure}
\includegraphics[width=\columnwidth]{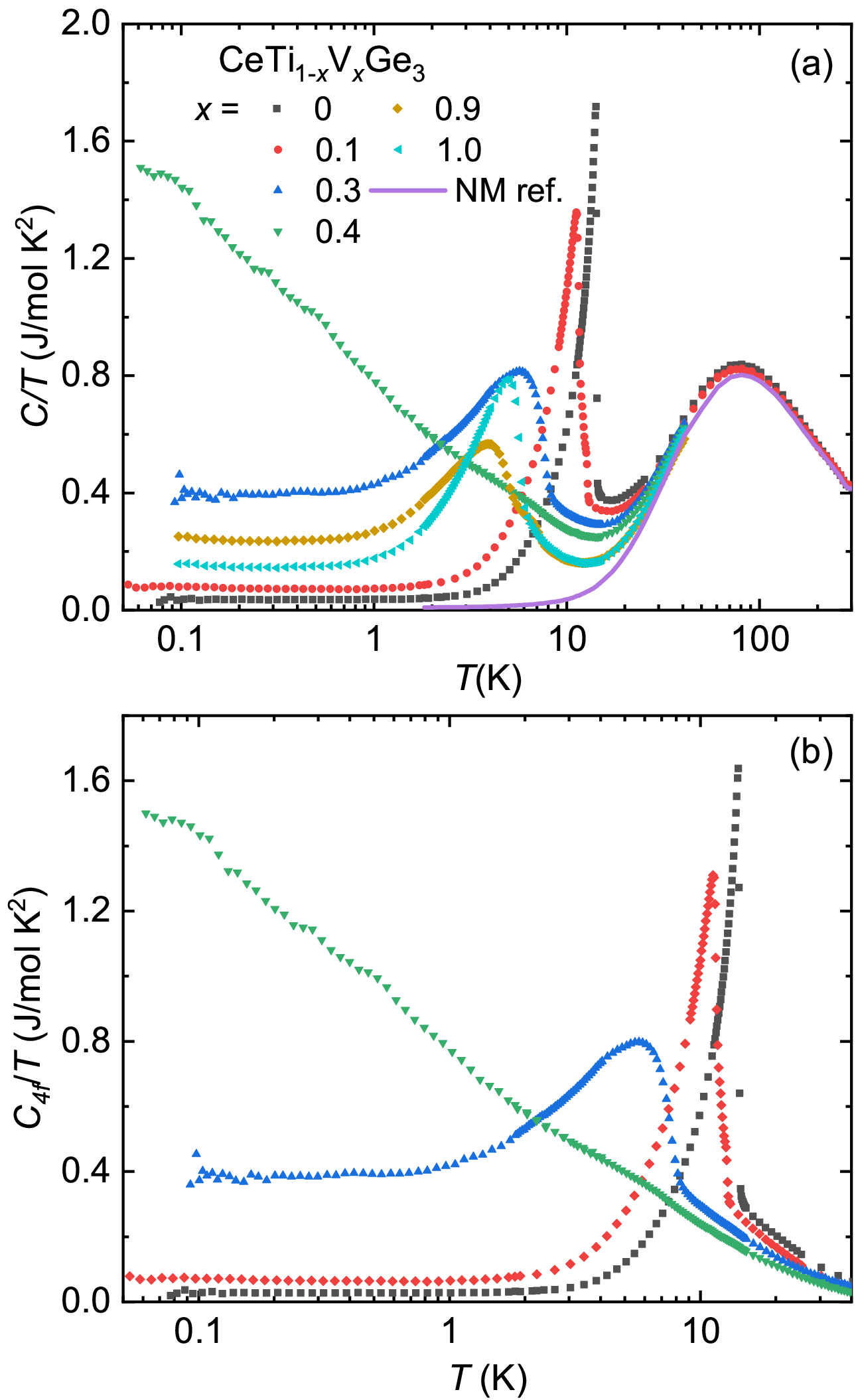}
\caption{(a) Zero-field specific heat of CTVG. NM ref. stands for nonmagnetic reference LaTi$_{0.55}$V$_{0.45}$Ge$_{3}$. (b) 4$f$ contributions to the specific heat.}  
\label{Fig1} 
\end{figure}

Figure~\ref{Fig1}(a) shows the total specific heat of CTVG and nonmagnetic LaTi$_{0.55}$V$_{0.45}$Ge$_{3}$. As $x$ increases from 0 to 0.3, the sharp FM transition around $T = 14$~K is suppressed to 7~K. As we approach lower temperatures, constant $C/T$ is observed for $x = 0-0.3$, indicating Fermi-liquid (FL) behavior. For the $x$ = 0.4, no transition could be observed down to 0.06~K, indicating that $x = 0.4$ is very close to a magnetic instability, i.e., the critical concentration $x_{cr}$ at which the FM transition disappears. As we further increase the V concentration to $x$ = 0.9 and 1.0, transitions appear around $T = 5-6$ K, below which a constant $C/T$ indicating the FL behavior is observed. Below the transition temperatures for samples with $x = 0.9$ and 1.0, an AFM order was reported~\cite{Jin2022}, and a helimagnetic state characterized by moments rotating around the $c$-axis has recently been fully elucidated~\cite{Chaffey2023}. 

We next focus solely on the FM side of CTVG ($x = 0-0.4$) and discuss the magnetic contribution to the specific heat. $C/T$ of the nonmagnetic reference is subtracted from that of CTVG, and the result is shown in Fig.~\ref{Fig1}(b). For $x = 0-0.3$, $C_{4f}/T$ exhibit a logarithmic increase as decreasing $T$ from high temperature to right above respective $T_{\rm C}$, and for $x =0.4$ the divergence of $C_{4f}/T$ extends widely between $T = 40$ and 0.06~K. Such an increase of $C_{4f}/T$ at temperature above 10~K can be described by the CEF splittings of the Ce$^{3+}$ ground state doublet, as evidenced by the comparison of the experimental data and the simulated CEF curves for $x = 0-0.3$ in Fig. S2(a-c) in the SM \cite{SM}. For $x =0.4$, we have analyzed the CEF splittings based on the temperature dependence of the inverse magnetic susceptibility derived from the same piece of the sample for the specific heat measurement, and the constraint for the high temperature magnetic entropy. We found the first excited state $\Delta_{1} =$ 12~K and the second excited state $\Delta_{2} =$ 65~K (see Fig. S2(d) in Ref.~\cite{SM}). Between $T = 10-40$~K, $C_{4f}/T$ matches $C_{\rm {CEF}}/T$ quite well. Below 10~K, $C_{4f}/T$ cannot be explained by the $C_{\rm {CEF}}/T$. Instead, we observe non-Fermi-liquid (NFL) behavior, i.e., logarithmic divergence $C_{4f}/T = a$ln$T/T_{0}$ with $a \sim -0.25$~J mol$^{-1}$ K$^{-2}$ and $T_{0} \sim 26$~K, between 0.06 and 10~K, i.e., over more than two orders of magnitude in $T$. The magnetic entropy $S_{4f}(T)$ obtained from $C_{4f}$ is shown in Fig.S1 in SM \cite{SM}. $S_{4f}$ exceeds $R$ln2 = 5.76 J/mol K at 16~K expected for a ground-state doublet. This is again due to the relatively small crystal-electric-field (CEF) splittings. If we express $C_{4f}/T$ as $\gamma + C_{\rm {CEF}}/T$, where $\gamma$ is the Sommerfeld coefficient, the observation of logarithmic divergence in $C_{4f}/T$ between 0.06 and 10~K even with the presence of the low CEF splitting implies the strong hybridization
between the conduction and $f$ electron states, i.e., $\gamma$ is dominant in $C_{4f}/T$ \cite{Konic2021}.

\begin{figure}
\includegraphics[width=\columnwidth]{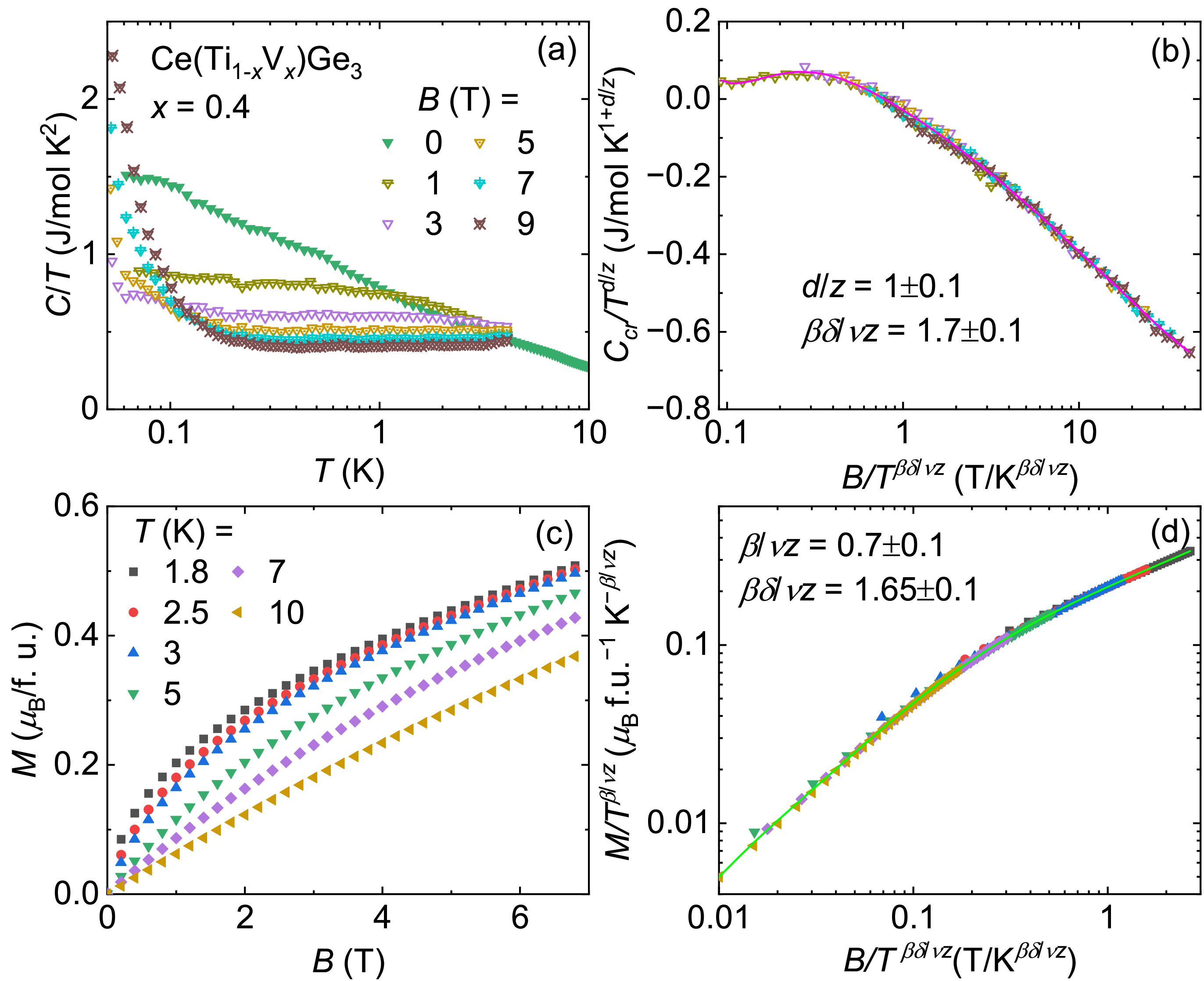}
\caption{(a) Specific heat of CTVG with $x = 0.4$ under different magnetic fields. (b) Scaling of $C_{cr}$ as a function of $T$ and $B$. (c) Isothermal magnetization curves measured at different temperatures. (d) Scaling of magnetization $M$ as a function of $T$ and $B$. The solid lines in (b) and (d) represent fits of polynomials to test the quality of the scaling.}  
\label{Fig2} 
\end{figure}

A straightforward way to qualitatively explain the NFL behavior at a magnetic instability is to consider the presence of numerous low-energy magnetic excitations when the $T_{c}\rightarrow 0$ \cite{Loehneysen1996}. This conjecture is supported by the recovery of FL behavior in high magnetic fields. Figure~\ref{Fig2}(a) shows the magnetic field dependence of $C/T$ of the $x = 0.4$ sample. After subtraction of the nuclear contribution to the specific heat $C_{nuclear}$ which causes an upturn at the lowest temperature, Figure S3 in the supplemental material shows the tendency towards FL behavior $(C-C_{nuclear})/T$ = constant is strengthened as the field increases \cite{SM}. 

Due to the insensitivity to low CEF splittings in specific heat, we are prompted to conduct further studies on the quantum criticality in CTVG. At a QCP below the upper critical dimension, the hyperscaling relation of the quantum critical part of the specific heat $C_{cr}$ is given by  
\begin{equation}
    \frac{C_{cr}(T,B)}{T^{d/z}}=\Psi\left (\frac{B}{T^{\beta\delta/\nu z}} \right ),
    \label{C_{cr}1}
\end{equation}
where $d$ is the spatial dimension, and $z$ and $\beta\delta/\nu$ are critical exponents associated with the tuning parameters $T$ and $B$, respectively \cite{Loehneysen2007,Lin2022}. Equation~\ref{C_{cr}1} indicates that $C_{cr}/T$ is a universal function of $B/T^{\beta\delta/\nu z}$. To eliminate non-critical quasiparticle contribution to the electronic specific heat, we determine $C_{cr} = C(T,B)-C(T,0)$ ($C_{nuclear}$ subtracted) and plot $C_{cr}/T^{d /z}$ vs. $B/T^{\beta\delta/\nu z}$. Excellent scaling over more than two orders of magnitude of $B/T^{\beta\delta/\nu z}$ with $d/z = 1\pm0.1$ and $\beta\delta/\nu z = 1.7\pm0.1$ is shown in Fig.~\ref{Fig2}(b) (also see Fig. S2 in \cite{SM}). As $C$ is tightly bind to the Gibbs free energy $\mathcal{F}$, from which the isothermal magnetization $M$ can be obtained, it is expected $M$ follows similar scaling behavior near a QCP, i.e., 
\begin{equation}
    \frac{M(T,B)}{T^{\beta/\nu z}}=\Phi \left( \frac{B}{T^{\beta\delta/\nu z}} \right ).
     \label{C_{cr}2}
\end{equation}
Figure~\ref{Fig2}(c) shows $M(B)$ measured at different temperatures. Using the Eq.~\ref{C_{cr}2}, excellent scaling over more than two orders of magnitude of $B/T^{\beta\delta/\nu z}$ with $\beta/\nu z = 0.7\pm0.1$ and $\beta\delta/\nu z = 1.65\pm0.1$ is obtained and shown in Fig.~\ref{Fig2}(d)(also see Fig. S2 in SM \cite{SM}). Below 1~T, non-critical contributions to $M$ cause deviations from scaling behavior, and hence the data are omitted. While the exact forms of $\Psi$ and $\Phi$ in Eqs.~\ref{C_{cr}1}  and~\ref{C_{cr}2}, respectively, determined by the details of $\mathcal{F}$ are not clear at the present stage, the scaling results still strongly suggest the existence of an FM QCP in CTVG with $x = 0.4$.    

\begin{figure}
\includegraphics[width=\columnwidth]{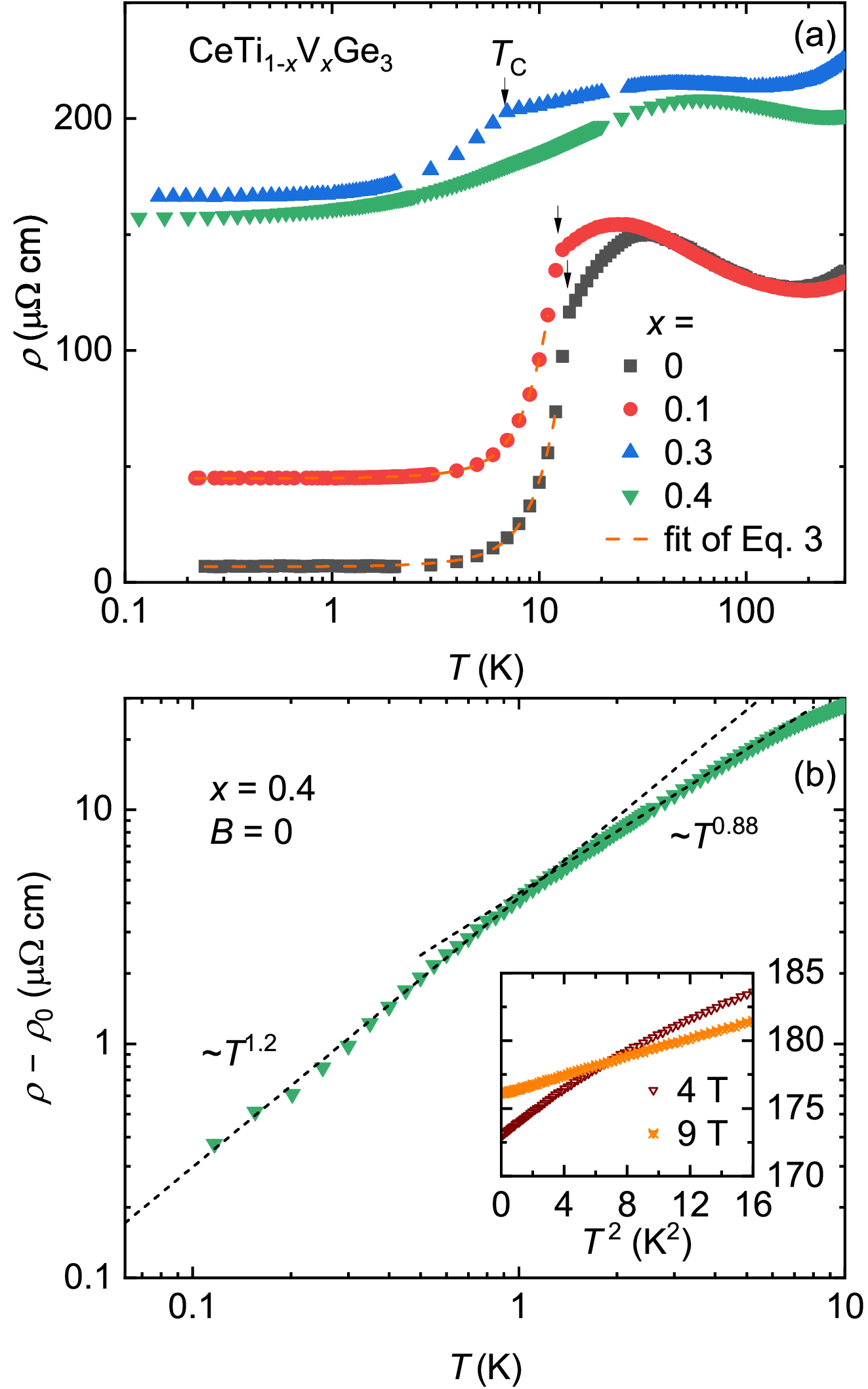}
\caption{(a) Zero-field electrical resistivity of CeTi$_{1-x}$V$_{x}$Ge$_{3}$ with $x = 0, 0.1, 0.3$ and 0.4. Arrows indicate the FM ordering temperature. (b) Low-temperature electrical resistivity of $x = 0.4$, obeying $\rho-\rho_{0} = AT^{\alpha}$, with $\rho_{0} = 156.85$ $\mu\ohm$ cm, $\alpha \sim 1.2$ between 1 and 0.1~K, and $\alpha \sim 0.88$ between 5 and 1~K. The inset shows $\rho(T)$ vs. $T^{2}$ for $B = 4$ and 9~T.}  
\label{Fig3} 
\end{figure}

In addition to the NFL behavior in the thermodynamic properties, at low temperatures, the scattering caused by quantum fluctuations drives quasiparticles critical, and this phenomenon is best represented by electrical resistivity measurements, as shown in Fig.~\ref{Fig3}. For $x = 0$ and 0.1, below the FM transition the $\rho(T)$ data can be described by 
\begin{equation}
   \rho =\rho_{0} + AT^{2} + BT\Delta(1+2\frac{T}{\Delta})e^{-\frac{\Delta}{T}}
     \label{rho}
\end{equation}
where $\rho_{0}$ is the residual resistivity, $A$ is the coefficient for
electron-electron scattering, $B$ is the coefficient for electron-magnon scattering, and $\Delta$ is an energy gap in the magnon excitation spectrum \cite{Jin2022}. The fits of Eq.~\ref{rho} are shown as dashed lines in Fig.~\ref{Fig3}(a). For $x = 0.3$ and 0.4, the Eq.~\ref{rho} failed to fit the low-$T$ data, indicating a complicated evolution of
the interplay between the Kondo effect, the CEF effect, and the quantum fluctuations. The low-$T$ part of $\rho(T)$ for $x = 0.4$ is shown in Fig.~\ref{Fig3}(b). The data shows appreciable NFL behavior, i.e., $\rho(T) = \rho_{0}+aT^{\alpha}$ with $\alpha \sim 1.2$, in the temperature range between 1 K and 0.1~K. The power $\alpha$ evolves to $\sim 0.88$ between 5 K and 1~K, which may be due to the coupling of both the CEF splitting and the Kondo effect to the ferromagnetic fluctuations. If the NFL behavior in $\rho(T)$ originates from FM fluctuations, an applied field should suppress low-lying magnetic excitations, and hence FL behavior prevails. The inset of Fig.~\ref{Fig3}(b) shows that $B = 9$~T leads to a recovery of a $\rho(T) = \rho_{0}+aT^{2}$ law. Our $\rho(T)$ data reinforces the evidence of the FM QCP in CTVG with $x = 0.4$. 

The source of NFL behavior within the CTVG system, as previously explained, can be linked to its proximity to a magnetic instability occurring at $x = 0.4$ at absolute zero. In some doped systems, the NFL behavior could be attributed to a distribution of the Kondo temperature $T_{\rm K}$ originated from inhomogeneities \cite{Bernal1995}. Such an effect could be neglected in CTVG as the NFL behavior only exists at the $x = 0.4$ sample. 

Next we discuss the nature of quantum criticality in CTVG. In 4$f$ heavy fermion systems upon increasing tuning parameters, the competition between the RKKY coupling among the moments and the Kondo interaction varies. Two classes of QCPs have been identified \cite{Gegenwart2008}: the Kondo-breakdown QCP, where the critical fluctuations encompass both the fluctuations of the magnetic order parameter and the breakdown of the Kondo effect, and the spin-density-wave (SDW) QCP, where critical fluctuations involve only fluctuations of the magnetic order parameter. In the Kondo-breakdown scenario, the NFL behavior is commonly observed over a wide "quantum critical fan" above the QCP \cite{Custers2003,Shen2020}, while in the SDW scenario the NFL behavior is largely limited to a small region above the QCP. In CTVG, the NFL behavior only exists at $x = 0.4$, which suggests the SDW criticality can better describe our system.
 
For a clean ferromagnet, the SDW-type QCP proposed originally by Hertz \cite{Hertz1976} is inherently unstable so that the QCP is disrupted by competing phases or first order phase transitions as $T_{\rm C}\rightarrow 0$ \cite{Belitz2005,Manuel2016}. The BKV theory suggests it is only possible to stabilize an FM QCP by introducing disorder. For CTVG, the fact that the ferromagnetism is destroyed by increasing the vanadium concentration could logically classify the QCP in CTVG within the framework of the BKV theory. Our hyperscaling of $C_{cr}(T,B)$ gives rise to the exponent $d/z = 1\pm0.1$ (Fig.~\ref{Fig2}(b)). The ratio of the 2-K magnetic susceptibility with the magnetic field applied parallel and perpendicular to the $c$-axis is $\sim 1$ for $x = 0.4$, suggesting $d =3$, as shown in Fig. S4 \cite{SM}. Therefore, $z = 3$ for CTVG, which is consistent with the prediction of the BKV theory in the dirty limit ($z = d$)\cite{Kirkpatrick2015}. The critical exponent $\delta = 2.4$ deduced from the hyperscaling of $M(T,B)$ (Fig.~\ref{Fig2}(d)) is slightly larger than the value predicted by the BKV theory ($\delta = 1.8$), which implies CTVG is in the preasymptotic region, i.e., strong disorder. Given the fact that the $\rho_{0}$ amounts to $\sim$ 157 $\mu\ohm$ cm for $x = 0.4$, much larger than the values for $x = 0$ and 0.1, our system indeed falls in the strong-disorder regime of the BKV theory. Moreover, the resistivity data shows $\Delta\rho \propto T^{1.2}$ at the QCP in CTVG, as shown in Fig.~\ref{Fig3}(b). The power of 1.2 is the value one expects for Hertz's fixed point in the presence of strong disorder \cite{Kirkpatrick2015}. We notice that the NFL behavior in $C_{4f}/T$ and $\rho$ is observed in different temperature windows. Similar phenomenon has been observed in an AFM QCP YbRh$_{2}$(Si$_{0.95}$Ge$_{0.05}$)$_{2}$, where quasiparticles in the quantum critical region may break up into spin (spinon) and charge (holon) excitations, dominating in different temperature regions for different measurements \cite{Custers2003,Custers2010}. Whether this scenario applies here needs further investigation.

To conclude, CTVG has revealed clear evidence of the FM QCP where the NFL behavior is registered in temperature dependence of $C_{4f}/T$ and $\rho$, as well as the existence of the hyperscaling on $M$ and $C_{cr}$. The obtained critical exponents and the temperature dependence of $\rho$ both agree well with the BKV theory in the strong-disorder regime. In the future, inelastic neutron scattering measurements on CTVG will be indispensable which allow us to quantitatively resolve the CEF splittings, especially for the samples in the vicinity of the FM QCP. This will help individually clarify the roles of different Ce$^{3+}$ doublets in terms of the quantum criticality.

\section*{Acknowledgements}

We thank Dr. M.-K. Lee at PPMS-16T Lab, Instrumentation Center, National Cheng Kung University (NCKU), and Ms. J. Kang at Center for Condensed Matter Sciences (CCMS), National Taiwan University (NTU) for technical support. We acknowledge D. Belitz for providing valuable comments and for reviewing the paper. This work is supported by the National Science and Technology Council in Taiwan (grant numbers NSTC 109-2112-M-006 -026 -MY3 and 112-2124-M-006-011) and the Higher Education Sprout Project, Ministry of Education to the Headquarters of University Advancement at NCKU. H.J, P.K. and V.T. acknowledge funding from the UC Laboratory Fees Research Program (LFR-20-653926). W.T.C. acknowledges for the supports from National Science and Technology Council in Taiwan (grant numbers NSTC 111-2112-M-002-044-MY3 and 112-2124-M-002-012), the Featured Areas Research Center Program at National Taiwan University (112L900802), and Academia Sinica project number AS-iMATE-111-12.

%
\end{document}